\documentclass{iauc}
\usepackage{graphicx}

\title[dEs and the dI Fundamental Plane] 
{dEs and the dI Fundamental Plane}

\author[Vaduvescu \& McCall]   
{Ovidiu Vaduvescu, Marshall L. McCall}

\affiliation{Department of Physics and Astronomy, York University \break
  128 Petrie Science Building, 4700 Keele St., Toronto, Ontario M3J 1P3, Canada \break
  email: ovidiuv@yorku.ca, mccall@yorku.ca\\[\affilskip]}

\pubyear{2005}
\volume{198}  
\pagerange{xxx--xxx}
\date{?? and in revised form ??}
\jname{IAU Colloquium 198, Les Diablerets, Switzerland}
\editors{H. Jerjen \& B. Binggeli, eds.}
\begin{document}

\maketitle

\begin{abstract}
Despite much work, the connection between dwarf elliptical galaxies (dEs) and 
dwarf irregulars (dIs) remains unclear. Recently, we found that the surface brightness 
profiles (SBPs) of dIs in the near-infrared can be well fitted with a hyperbolic 
secant (sech) function. From sech fits, a tight relationship was derived between the 
absolute magnitude, central surface brightness, and 21cm line widths, which amounts 
to a fundamental plane for dIs. Here we examine how closely dEs fit into the dI 
fundamental plane using published data for 22 dEs in the Virgo cluster and the 
Local Group. Over a 9 mag interval in absolute magnitude the dEs fall in the plane 
defined by the dIs. The outstanding overlap suggests more than a casual relationship 
between the two classes. 
\end{abstract}

\section{The dI Sample}
A sample of 19 field dIs with very accurate distances was observed at CFHT 
adopting a prudent approach to sample the faint outer regions 
(\cite[Vaduvescu and McCall 2004]{vaduvescu04}). 
We separated resolved and unresolved 
components using the KILLALL package (\cite[Buta and McCall 1999]{buta}) and then 
derived the surface brightness profiles (SBPs) by fitting ellipses using the 
stsdas/ellipse function in IRAF. The hyperbolic secant function (sech) was found to 
fit accurately all the SBPs in our dI sample (\cite[Vaduvescu \etal\ 2005]{vaduvescu05}): 
\begin{equation}
I = I_0 \hbox{ sech} {(r/r_0)} = \frac{2I_0}{e^{r/r_0}+e^{-r/r_0}}
\end{equation}
\noindent
where $I_0$ is the central surface brightness and $r_0$ the scale length. Note that 
the sech function becomes exponential at large radius. 

\section{The dE Sample}

The sample of dEs consists of 22 objects with published data in three sub-samples: 
11 dEs in the Virgo Cluster with H band images from the GOLDMine database 
(\cite[Gavazzi \etal\ 2002]{gavazzi}) and velocity dispersions from 
\cite[Geha \etal\ (2003)]{geha}; three dEs/dSphs in the Local Group 
(M 110, NGC 147, NGC 185) with K images from 2MASS Large Galaxy Atlas 
(\cite[Jarett \etal\ 2003]{jarett}) and velocity dispersions from 
\cite[McElroy (1995)]{mcelroy} and \cite[Held \etal\ (1992)]{held}; 
8 dSphs orbiting our Galaxy (Carina, Draco, Fornax, Leo I, Leo II, Sculptor, 
Sextans, Ursa Minor) for which star counts in V define surface brightness 
profiles (\cite[Irwin and Hadzidimitriou 1995]{irwin}). We adopted average 
colors of H-K=0.2 mag and V-K=2.7 mag. In all cases, an exponential model 
was adopted to fit the outer isophotes: $I = I_0 e^{-r/r_0}$. 

\section{The Tully-Fisher Relation and the Fundamental Plane}

In Figure~1 (left), we plot the Tully-Fisher relation for both the dIs and dEs 
with $W_{20}$ estimated for dEs from stellar velocity dispersions. It shows 
considerable scatter. 

In Figure~1 (right), we show the fundamental plane defined by the absolute 
magnitude $M_K$, the central surface brightness $m_0$ derived from $I_0$, and the 
HI line-width $W_{20}$. For the dIs, 
the ``absolute magnitude'' $M_K$ is the sech magnitude, while for dEs it is the 
exponential magnitude. The dIs and dEs are marked by crosses and circles, 
respectively. No corrections for tilt have been applied. The dotted line 
represents the fit to the dIs alone (defined by the abscissa). The dI fundamental 
plane shows a remarkable good correlation with a coefficient of correlation 0.91 
and a rms scatter of only 0.4 mag. This is comparable to the scatter in the I-band 
Tully-Fisher relation for spirals adopted in the HST Key Project on the Extragalactic 
Distance Scale. Over a 9 mag interval in absolute magnitude, the dEs fall in the dI 
fundamental plane. The outstanding overlap suggests a common scenarion for the 
early development of the dEs and dIs. 

\begin{figure}[h]
\centering
\includegraphics[width=6cm]{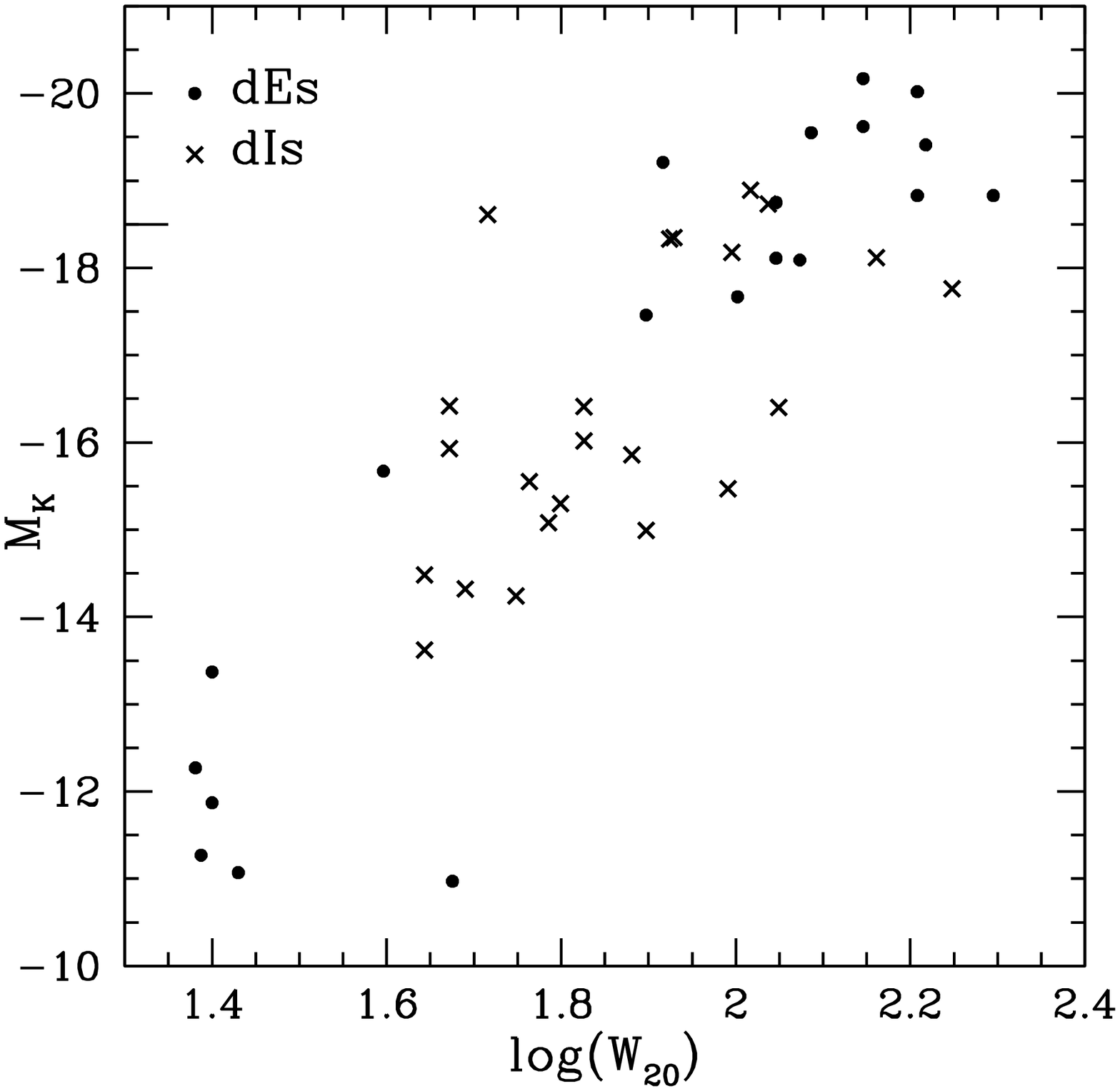}
\includegraphics[width=6cm]{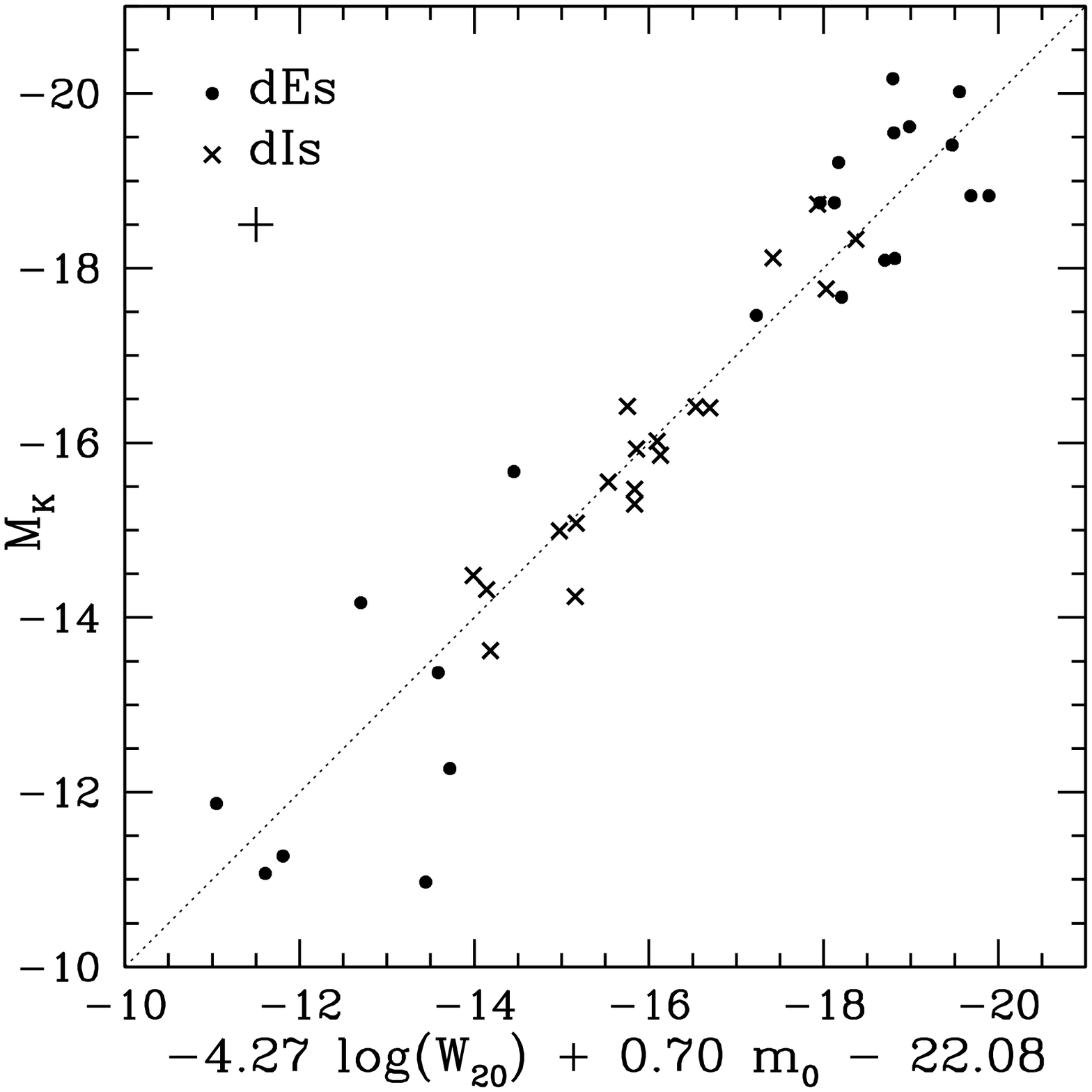}
  \caption{Left: The Tully-Fisher relation appears to hold for both dIs and dEs, 
   although with much scatter. Right: The Fundamental Plane obtained by fitting 
   HI line-widths $W_{20}$, central brightness $m_0$, and the sech absolute 
   magnitude $M_K$ for 19 dIs with available data 
   (\cite[Vaduvescu \etal\ 2005]{vaduvescu05}). The dEs lie on the same plane. 
   A common distance modulus DM=30.62 has been applied for Virgo members. 
   Typical uncertainties are plotted as an error cross. }
\end{figure}

\begin{acknowledgments}
We are grateful to the Natural Sciences and Engineering Research Council of Canada 
for its continuing support to the PhD Thesis of OV (2005, in preparation). 
This research has made use of the GOLDMine Database (\cite[Gavazzi \etal\ 2002]{gavazzi}) 
and data products from the Two Micron All Sky Survey. OV would like to thank to Stefano 
Zibetti for some feedback related to data reduction of the GOLDMine images. 
\end{acknowledgments}

\end{document}